\documentclass[pdflatex,prl,twocolumn,showpacs,superscriptaddress,nofootinbib]{revtex4-1}
\usepackage{bm,amsmath,amssymb}
\usepackage{graphicx}
\usepackage{subfigure}
\usepackage{color}
\usepackage{soul}
\usepackage{hyperref}
\usepackage{multirow}
\usepackage{lineno}
\usepackage{footnote}
\setulcolor{blue}
\setstcolor{red}
\sethlcolor{yellow}

\newcommand \beq {\begin{equation}}
\newcommand \eeq {\end{equation}}
\newcommand \beqa {\begin{eqnarray}}
\newcommand \eeqa {\end{eqnarray}}

\def\lsim{\raise0.3ex\hbox{$<$\kern-0.75em\raise-1.1ex\hbox{$\sim$}}}
\def\gsim{\raise0.3ex\hbox{$>$\kern-0.75em\raise-1.1ex\hbox{$\sim$}}}
\begin{document}

\title{Chiral phase transition temperature in (2+1)-Flavor QCD}

\author{H.-T. Ding}
\affiliation{ Key Laboratory of Quark \& Lepton Physics (MOE) and Institute of
Particle Physics, Central China Normal University, Wuhan 430079, China}
\author{P. Hegde}
\affiliation{Center for High Energy Physics, Indian Institute of Science,
Bangalore 560012, India}
\author{O. Kaczmarek}
\affiliation{ Key Laboratory of Quark \& Lepton Physics (MOE) and Institute of
Particle Physics, Central China Normal University, Wuhan 430079, China}
\affiliation{Fakult\"at f\"ur Physik, Universit\"at Bielefeld, D-33615 Bielefeld,
Germany}
\author{F. Karsch}
\affiliation{Fakult\"at f\"ur Physik, Universit\"at Bielefeld, D-33615 Bielefeld,
Germany}
\affiliation{Physics Department, Brookhaven National Laboratory, Upton, New York 11973, USA}
\author{Anirban Lahiri}
\affiliation{Fakult\"at f\"ur Physik, Universit\"at Bielefeld, D-33615 Bielefeld,
Germany}
\author{S.-T.~Li}
\affiliation{ Key Laboratory of Quark \& Lepton Physics (MOE) and Institute of
Particle Physics, Central China Normal University, Wuhan 430079, China}
\author{Swagato Mukherjee}
\affiliation{Physics Department, Brookhaven National Laboratory, Upton, New York 11973, USA}
\author{H. Ohno}
\affiliation{Center for Computational Sciences, University of Tsukuba, Tsukuba, Ibaraki 305-8577, Japan \\[2mm]
{\bf (\centering{HotQCD Collaboration)}
}}
	
\author{P. Petreczky}
\affiliation{Physics Department, Brookhaven National Laboratory, Upton, New York 11973, USA}
\author{C. Schmidt}
\affiliation{Fakult\"at f\"ur Physik, Universit\"at Bielefeld, D-33615 Bielefeld,
Germany}
\author{P. Steinbrecher}
\affiliation{Physics Department, Brookhaven National Laboratory, Upton, New York 11973, USA}

\begin{abstract}

We present a lattice-QCD-based determination of the chiral phase transition
temperature in QCD with two degenerate, massless quarks and a physical strange quark
mass using lattice QCD calculations with the Highly Improved Staggered Quarks
action. We propose and calculate two novel estimators for the chiral transition
temperature for several values of the light quark masses, corresponding to Goldstone
pion masses in the range of $58~{\rm MeV}\lsim m_\pi\lsim 163~{\rm MeV}$. The chiral
phase transition temperature is determined by extrapolating to vanishing pion
mass using universal scaling analysis.  Finite-volume effects are controlled by
extrapolating to the thermodynamic limit using spatial lattice extents in the range
of $2.8$-$4.5$ times the inverse of the pion mass. 
Continuum extrapolations are carried out by using
three different values of the lattice cutoff,  corresponding to lattices with
temporal extents $N_\tau=6,\ 8$ and $12$.  After thermodynamic, continuum, and chiral
extrapolations we find the chiral phase transition temperature
$T_c^0=132^{+3}_{-6}$~MeV.
\end{abstract}


\maketitle
\emph{Introduction.---} For physical values of the light up, down, and heavier strange
quark masses strongly interacting matter undergoes a transition from a low-temperature hadronic regime to a high-temperature region that is best described by
quark and gluon degrees of freedom. This smooth crossover between the two asymptotic
regimes is not a phase transition~\cite{Ding:2015ona}. It is characterized by a
pseudocritical temperature, $T_{pc}$, that has been determined in several numerical
studies of Quantum Chromodynamics
(QCD)~\cite{Aoki:2009sc,Bazavov:2011nk,Bonati:2015bha}. A recent determination of
$T_{pc}$ extracted from the maximal fluctuations of several chiral observables
gave $T_{pc}= (156.5\pm 1.5)$~MeV~\cite{Bazavov:2018mes}.

In the chiral limit of (2+1)-flavor QCD, i.e., where two (degenerate) light
quark masses $m_l=(m_u+m_d)/2$ approach zero but the strange quark mass $m_s$ is kept fixed to
its physical value,  the pseudocritical behavior is expected to give rise to a
``true" chiral phase transition \cite{Kogut:1982fn,Pisarski:1983ms}.
The chiral phase transition temperature itself is expected to set an
upper bound on the temperature at a possible critical point at nonzero
baryon chemical potential \cite{Halasz:1998qr,Karsch:2019mbv}, which 
 is intensively searched for in heavy ion collision experiments. Whether this
chiral phase transition is first or second order may depend crucially on the
temperature dependence of the chiral anomaly~\cite{Pisarski:1983ms}.  In the latter
case critical behavior generally is expected to be controlled by the 3-D $O(4)$
universality class, although a larger 3-D universality
class~\cite{Grahl:2013pba,Pelissetto:2013hqa} may become of relevance in case the
axial anomaly also gets restored effectively at $T_c^0$. 
If the chiral phase transition is first order, then a second-order phase transition, belonging to the 3-D $Z(2)$
universality class, would occur for $m_l^c>0$. When decreasing the light to strange
quark mass ratio, $H=m_l/m_s$, towards zero, this would give rise to diverging susceptibilities
already for some critical mass ratio $H_c=m_l^c/m_s >0$. The analysis presented here
leads to a determination of the critical temperature $T_c^{H_c}$. However, as we do
not have any evidence for $H_c\ne0$, we de facto present a determination
of the chiral phase transition temperature $T_c^0$.

Although $T_c^0$ appears as a fit parameter in all finite-temperature scaling studies
of the chiral transition in QCD \cite{Ejiri:2009ac,Bazavov:2011nk,Burger:2018fvb},
so far  no lattice QCD calculation has carried out a systematic analysis of $T_c^0$
by controlling thermodynamic, continuum and chiral limits. Here,  we will present
a first lattice-QCD-based determination of $T_c^0$ in $(2+1)$-flavor QCD with
controlled thermodynamic, continuum and chiral extrapolations.  QCD-inspired model
calculations~\cite{Berges:1997eu,Braun:2005fj} suggest that $T_c^0$ might be even
lower by ($20-30$~MeV) than  $T_{pc}$.  To mitigate this potentially large
$m_l$ dependence of $T_{pc}$ while approaching $m_l\to0$, we propose two novel
estimators of the pseudocritical temperature having only mild dependence on $m_l$,
leading to well-controlled chiral extrapolation.

\emph{Chiral observables.--} 
At low temperatures, chiral symmetry is spontaneously broken in QCD. An 
order parameter for the restoration of this symmetry at high temperature 
is the chiral condensate, which is obtained as the derivative of the 
partition function, $Z(T,V,m_u,m_d,m_s)$, with respect to one of the quark 
masses, $m_f$,
\begin{equation}
\langle \bar\psi \psi\rangle_f = \frac{T}{V} \frac{\partial 
\ln Z(T,V,m_u,m_d,m_s)}{\partial m_f} \; .
\label{pbp}
\end{equation}
The light quark chiral condensate, 
$\langle \bar\psi \psi\rangle_l= (\langle \bar\psi \psi\rangle_u+\langle \bar\psi \psi\rangle_d)/2$, 
is an order parameter for the chiral phase 
transition that occurs in the limit $m_l\rightarrow 0$. For nonvanishing $m_l$,
this order parameter requires additive and multiplicative
renormalization. We take care of this by introducing a combination of the light and strange quark chiral condensates,
\begin{equation}
M = 2 \left( m_s \langle \bar\psi \psi\rangle_l - m_l \langle \bar\psi \psi\rangle_s
\right)/f_K^4 \; ,
\label{M}
\end{equation}
where the kaon decay constant, 
$f_K=156.1(9)/\sqrt{2}$~MeV,
for physical values of the degenerate light and strange quark mass,
is used as a normalization constant to define a dimensionless order parameter $M$. The order parameter $M$ is free of UV divergences linear in the quark
masses $m$ \cite{Bazavov:2011nk} but may still receive divergent contributions proportional to $m^3 \ln(m)$ which we neglect here.
The derivative  of $M$ with respect to the light quark masses gives the chiral susceptibility, 
\begin{eqnarray}
	\hspace*{-0.2cm}\chi_M &=& \left. 
	m_s (\partial_{m_u}+\partial_{m_d}) M \right|_{m_u=m_d}
\nonumber \\
&=&m_s( m_s \chi_l - 2 \langle \bar\psi \psi\rangle_s -4 m_l \chi_{su}
)/f_K^4\; ,
\label{chim}
\end{eqnarray}
with 
$\chi_{fg}=\partial_{m_f}  \langle \bar\psi \psi\rangle_g$
and $\chi_l= 2(\chi_{uu}+\chi_{ud})$.

\begin{figure}[!t]
\includegraphics[scale=0.6]{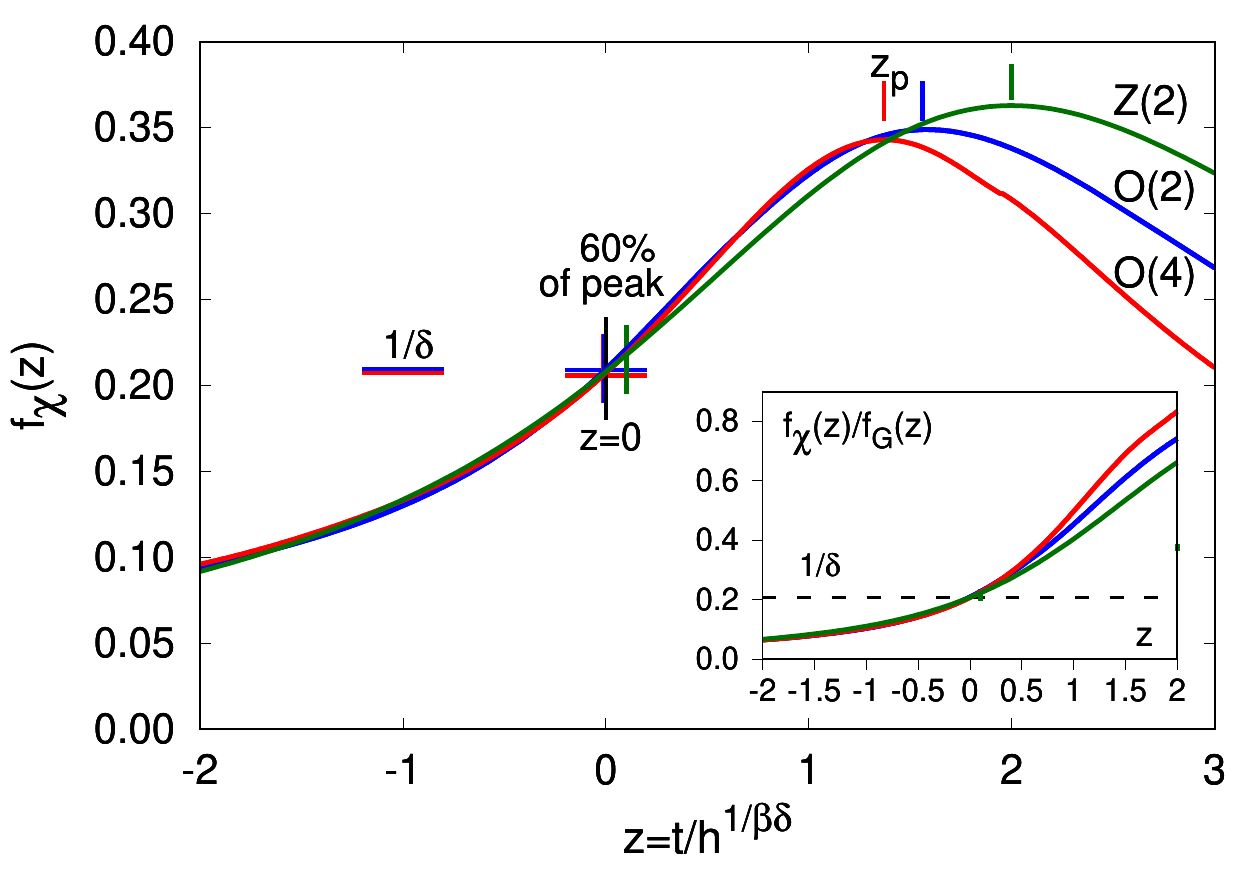}
\caption{Scaling functions for the 3-D $Z(2)$, $O(2)$ and $O(4)$
universality classes. The position $z_p$ of the peak of the scaling functions
(vertical lines) and the position $z_{60}$ where the scaling function attains
60\% of its maximal value (crosses) are also given in Table~\ref{tab:zpz60}.
Lines close to $z=-1$ show $1/\delta$ for these three universality classes, 
which agree to within better than 1\%. The inset shows the ratio of scaling functions,
$f_\chi(z)/f_G(z)$, used in determinations of the chiral phase transition
temperature.
}
\label{fig:scaling}
\end{figure}

\begin{table}[t]
\begin{center}
\begin{tabular}{|c|c|c|r||c|c|c|}
\hline
&$\delta$&$z_p$ & $z_{60}$~~~~~&$f_G(z_p)$ & $f_\chi (z_p)$& $r_\chi (0)$ \\
\hline
\hline
Z(2)&4.805& 2.00(5) &0.10(1)&0.548(10)&0.3629(1)& 0.573(1) \\
O(2)&4.780& 1.58(4) &-0.005(9)& 0.550(10)&0.3489(1)&0.600(1) \\
O(4)&4.824& 1.37(3) &-0.013(7)& 0.532(10)&0.3430(1)&0.604(1) \\
\hline
\end{tabular}
\end{center}
\caption{The critical exponent
$\delta$, location of the peak, $z_p$, and the position of 60\% of the peak value, 
$z_{60}$, of the scaling functions $f_\chi(z)$ for different 3-$d$ universality classes \cite{Engels:2002fi,Engels:2000xw,Engels:2011km}.
Also given are $f_G(z_p)$, $f_\chi(z_p)$ and $r_\chi(0) = f_\chi (0)/f_\chi (z_p)$.
}
\label{tab:zpz60}
\end{table}

When approaching the chiral limit, one also needs to control the
thermodynamic limit, $V \rightarrow \infty$. 
In the vicinity of a second-order
order phase transition,
$M$ and $\chi_M$ are given in
terms of the universal finite-size scaling functions $f_G(z,z_L)$ and
$f_\chi(z,z_L)$, which depend on the scaling variables 
$z=t/h^{1/\beta\delta}$ and $z_L= l_0/(Lh^{\nu/\beta\delta})$. Here
$t=(T-T_c^0)/(t_0 T_c^0)$ denotes the reduced temperature; $h=H/h_0$ is the
symmetry-breaking field; and $L/l_0$ parametrizes the finite size
of the system, $L\equiv V^{1/3}$. These scaling variables are expressed
in terms of nonuniversal parameters, $t_0,\ h_0,\ l_0$.

While the universal scaling functions control the behavior of $M$ and
$\chi_M$ close to a critical point at $(z,z_L)=(0,0)$, they 
also receive contributions from corrections to scaling and regular terms
\cite{Hasenbusch:2000ph,Engels:2000xw},
which we represent by a function $f_{\rm sub}(T,H,L)$. With this we may write
\begin{eqnarray}
M &=& h^{1/\delta} f_G(z,z_L) + f_{sub}(T,H,L) \; , \nonumber \\
\chi_M &=& h_0^{-1} h^{1/\delta-1} f_\chi(z,z_L) +\tilde{f}_{sub}(T,H,L) \; .
\label{scale}
\end{eqnarray}
As far as is needed for the analysis, we will specify contributions arising
from $f_{\rm sub}(T,H,L)$ later.

Close to the thermodynamic limit,
$f_\chi(z,z_L)$ has a pronounced peak, which often is used to define
a pseudocritical temperature, $T_p$. In the scaling regime this peak is 
located 
at some $z=z_p(z_L)$, which defines $T_p$,
\begin{equation}
T_p(H,L)= 
T_c^{0} \left( 1+ \frac{z_p(z_L)}{z_0} H^{1/\beta\delta} \right)\ +\ {\rm sub~leading} \; ,
\label{Tpc}
\end{equation}
with $z_0=h_0^{1/\beta\delta}/t_0$.
While the first term describes the universal quark mass dependence of $T_p$,
corrections may arise from
corrections to scaling and regular terms, shifting the peak location of
the chiral susceptibilities.

\begin{figure*}[!t]
\includegraphics[width=0.45\textwidth]{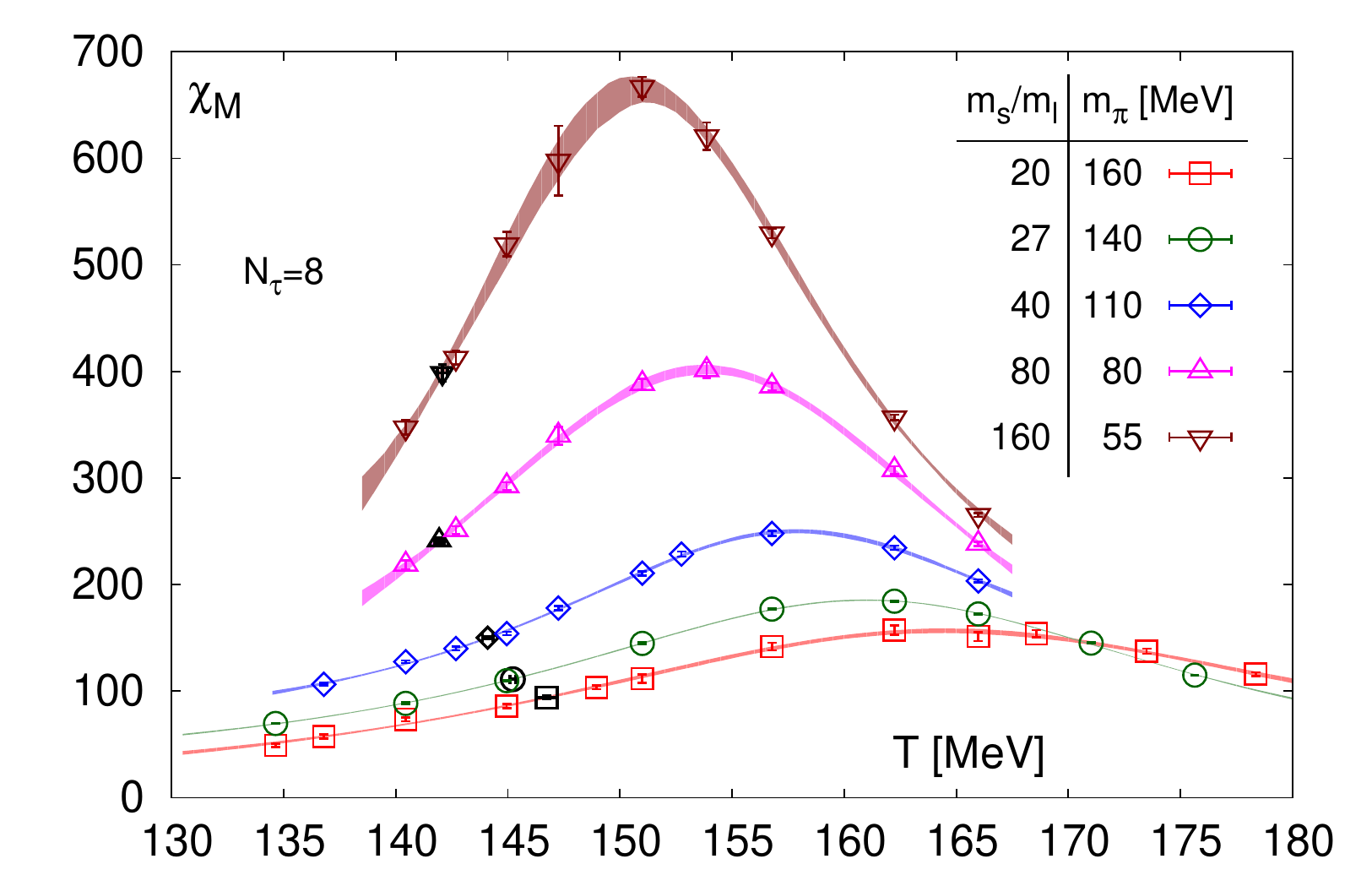}
\includegraphics[width=0.45\textwidth]{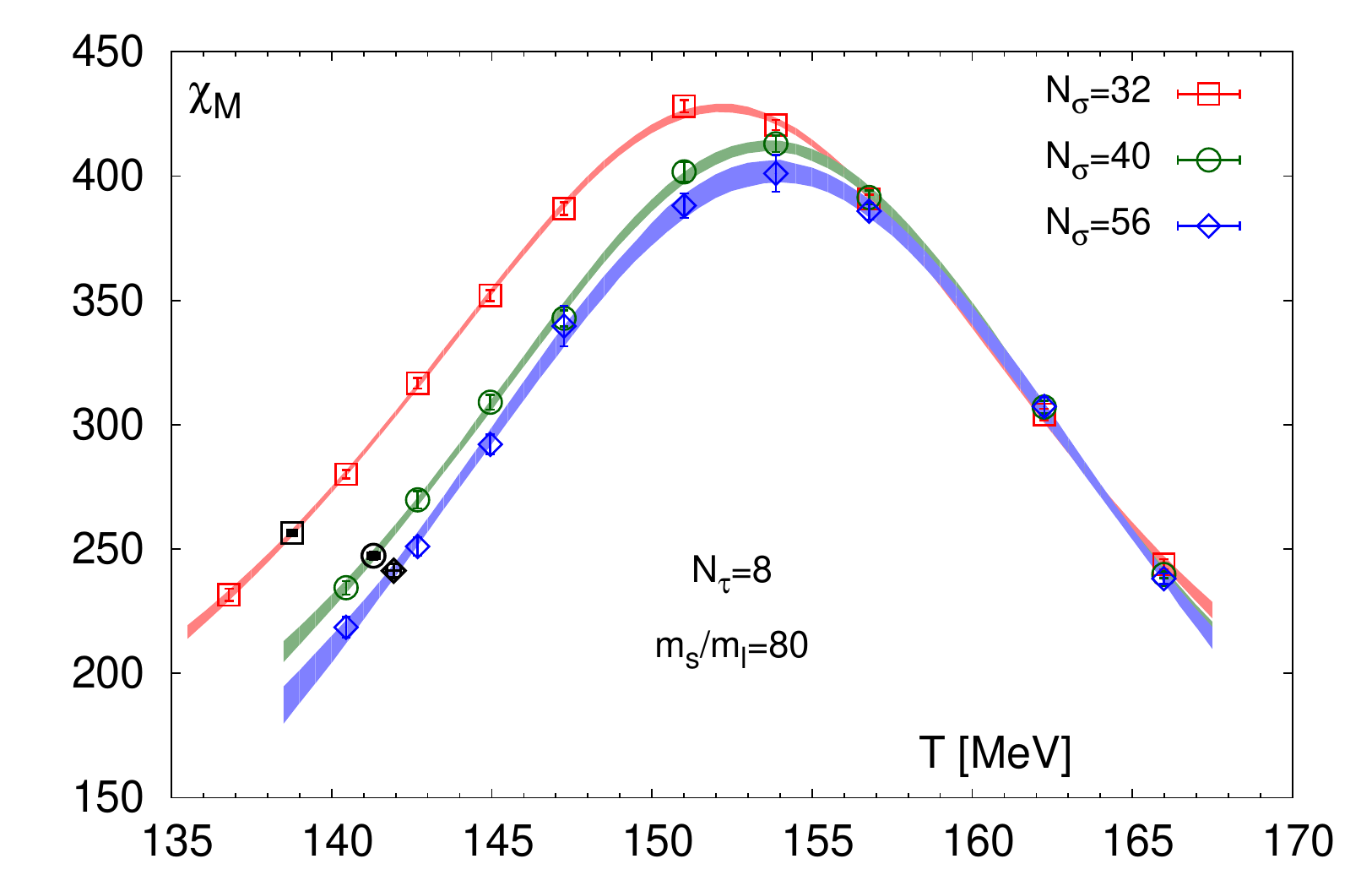}
\caption{Quark mass (left) and volume (right) dependence of
the chiral susceptibility on lattices with temporal extent
$N_\tau=8$. The left-hand figure shows
results for several values of the quark masses.
The spatial lattice extent $N_\sigma$ is
increased as the light quark mass decreases: $N_\sigma= 32$~
$(H^{-1}=20,\ 27)$, 40~$(H^{-1}=40)$, 56~$(H^{-1}=80,\ 160)$.
The right-hand figure shows results for three different spatial
lattice sizes at $H=1/80$. Black symbols mark the points corresponding
to 60\% of the peak height. 
}
\label{fig:sus}
\end{figure*}

When approaching the chiral limit, depending on the magnitude of
$z_p/z_0\equiv z_p(0)/z_0$,  $T_p(H,L)$
may change significantly with $H$. In the potentially large temperature interval
between  $T_c^{0}$ and $T_p(H,L)$, regular contributions, arising from
$f_{\rm sub}(T,H,L)$, may also be large, and during the $H\rightarrow 0$ extrapolation
several nonuniversal parameters may be needed to account for contributions 
from $f_{\rm sub}(T,H,L)$.~It is thus advantageous to determine $T_c^{0}$ using observables defined close to
$z\simeq0$. While $T_p(H,L)$, defined through such observables for small $H>0$, will have
milder $H$-dependence, the determination of $T_c^{0}=T_p(H\to0,L\to\infty)$ will be well controlled.

We will consider here two estimators for $T_c^0$, defined at or close to $z=0$. 
We determine temperatures $T_\delta$ and $T_{60}$ by demanding
\begin{eqnarray}
\frac{H \chi_M (T_\delta,H,L)}{M(T_\delta,H,L)} &=&\frac{1}{\delta}  \; ,
\label{ratio}\\
\chi_M(T_{60},H) &=& 0.6 \chi_M^{max} \; .
\label{T60}
\end{eqnarray}
Equation~\ref{ratio} has already been introduced in Ref.~\cite{Karsch:1994hm} 
as a tool to analyze the chiral transition in QCD,
and it is understood that $T_{60}$ is determined at a temperature on the
left of the peak $\chi_M^{\rm max}$, {\it i.e.}\ $T_{60}<T_p$.
These relations define pseudocritical temperatures, 
$T_X$, which are close to $T_c^0$ already for nonzero $H$ and $L^{-1}$. 
They converge to the chiral phase transition temperature $T_c^0$ in the 
thermodynamic and chiral limits. For nonzero $L^{-1}$, 
Eqs.~(\ref{ratio}) and (\ref{T60}) involve scaling variables $z_X(z_L)$ which
approach or are close to zero in the limit $L^{-1}\rightarrow 0$, i.e.,
$z_\delta\equiv z_\delta(0)=0$ and $z_{60}\equiv z_{60}(0) \simeq 0$.
Some values for $z_{60}$, for several universality classes, are
given in Table~\ref{tab:zpz60}, and the relevant scaling functions, obtained
in the thermodynamic limit $z_L=0$, are shown in Fig.~\ref{fig:scaling}.

Ignoring possible contributions from 
corrections to scaling, and keeping in $f_{\rm sub}$ only the leading 
$T$-independent, infinite-volume regular contribution proportional to $H$,
we then find for the pseudocritical temperatures
\begin{eqnarray}
T_X(H,L)&=& 
T_c^{0} \left( 1+ \left( \frac{z_X(z_L)}{z_0} \right) 
H^{1/\beta\delta} \right)
\nonumber \\
&&+c_X H^{1-1/\delta+1/\beta\delta}\;\; ,\;\; X=\delta,\ 60
 \; .
\label{TX}
\end{eqnarray}

The universal functions, $z_X (z_L)$ may directly be determined from the 
ratio of scaling functions, $f_\chi(z_\delta,z_L)/f_G(z_\delta,z_L)=1/\delta$
and $f_\chi(z_{60},z_L)/f_\chi(z_p,z_L)=0.6$, respectively. 
The finite-size scaling functions $f_G(z,z_L)$, $f_\chi(z,z_L)$ 
have been determined  for the 3-D $O(4)$ universality
class in Ref.~\cite{Engels:2014bra}. 

We will present here results on $T_\delta$ and $T_{60}$ obtained in lattice 
QCD calculations \cite{Ding:2018auz}.
We have calculated the chiral order parameter $M$ and the chiral
susceptibility $\chi_M$ [Eqs.~(\ref{M}) and (\ref{chim})]
in $(2+1)$-flavor QCD with degenerate up and down quark masses ($m_u=m_d$).
For our lattice QCD calculations, performed with the 
Highly Improved Staggered Quark (HISQ) action
\cite{Follana:2006rc} in the fermion sector and the Symanzik improved gluon action,
the strange quark mass has been tuned
to its physical value \cite{Bazavov:2014pvz}, and the
light quark mass has been varied in a range $m_l\in [m_s/160:m_s/20]$
corresponding to Goldstone
pion masses in the range $58~{\rm MeV}\lsim m_\pi \lsim 163~{\rm MeV}$.
At each temperature, we performed calculations on lattices of size
$N_\sigma^3 N_\tau$ for three different values of the lattice cutoff, 
$aT = 1/N_\tau$, with $N_\tau=6,\ 8$, and $12$.  In the HISQ discretization scheme, so-called taste symmetry violations
give rise to a distortion of the light pseudoscalar (pion) meson masses.
These discretization effects are commonly expressed in terms of a 
root-mean-square (RMS) pion mass which approaches the Goldstone pion mass in the continuum limit. For our computational setup and the three different values of the lattice cutoff this has been discussed in Ref. \cite{Bazavov:2011nk}. For lattice spacings corresponding to $N_\tau = 6,8$, and 12 one finds for physical values of the quark masses $M_{\rm RMS}=400$, $300$, and $200$~MeV,
respectively.
The spatial lattice extent, $N_\sigma=L/a$, has been varied in the range 
$4\le N_\sigma / N_\tau \le 8$.
For each $N_\tau$
we analyzed the volume dependence
of $M$ and $\chi_M$ in order to perform controlled infinite-volume
extrapolations. 

\emph{Results ---}
In Fig.~\ref{fig:sus}~(left) we show results for $\chi_M$ on lattices with
temporal extent $N_\tau=8$ for five different values of the quark mass ratio, 
$H=m_l/m_s$, and the largest lattice available for each $H$. 
The increase of the peak height, $\chi_M^{\rm max}$, with decreasing $H$ is 
apparent. This rise is consistent with the expected behavior,
$\chi_M^{\rm max} \sim H^{1/\delta-1}+ const.$, with  $\delta \simeq 4.8$;
however, a precise
determination of $\delta$ is not yet possible with the current data.

\begin{figure*}[!t]
\includegraphics[width=0.32\textwidth]{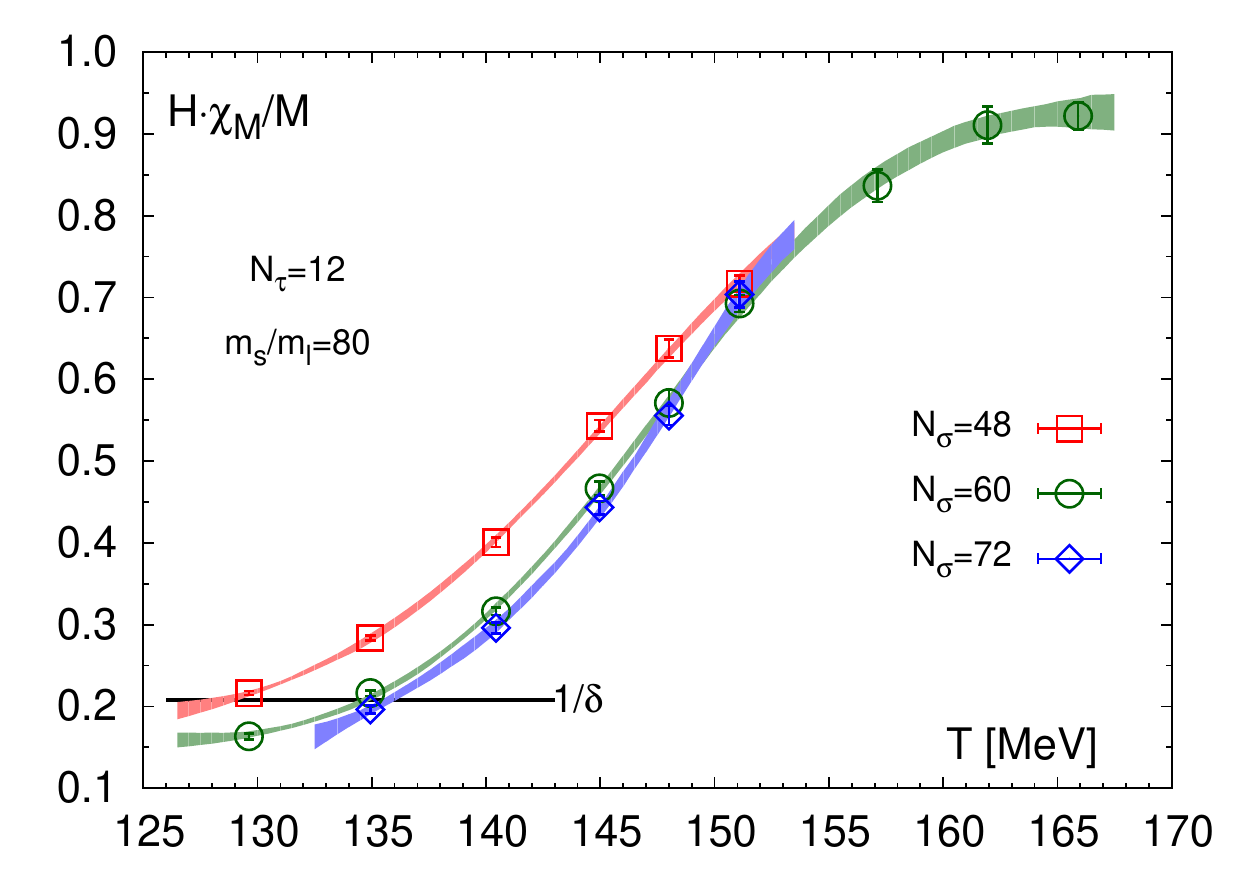}
\includegraphics[width=0.32\textwidth]{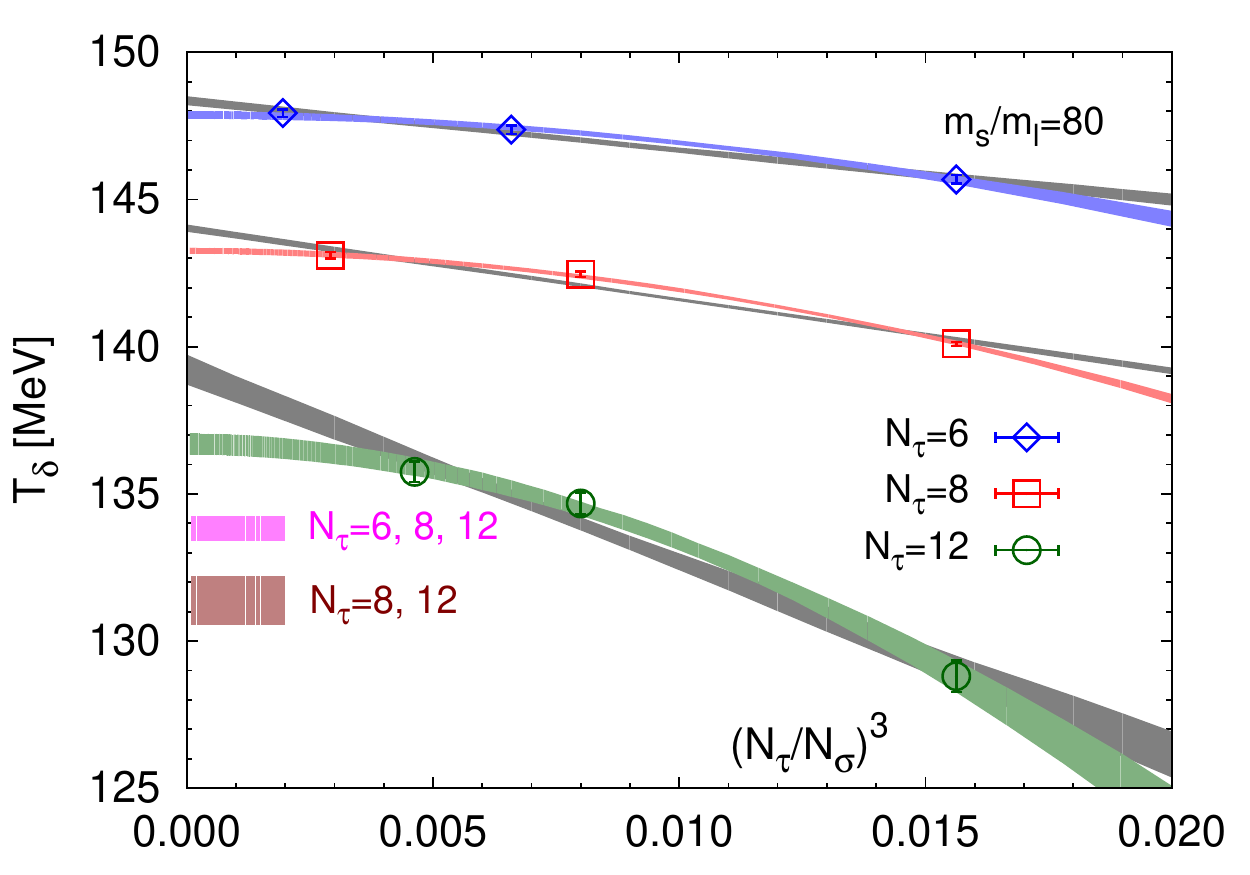}
\includegraphics[width=0.32\textwidth]{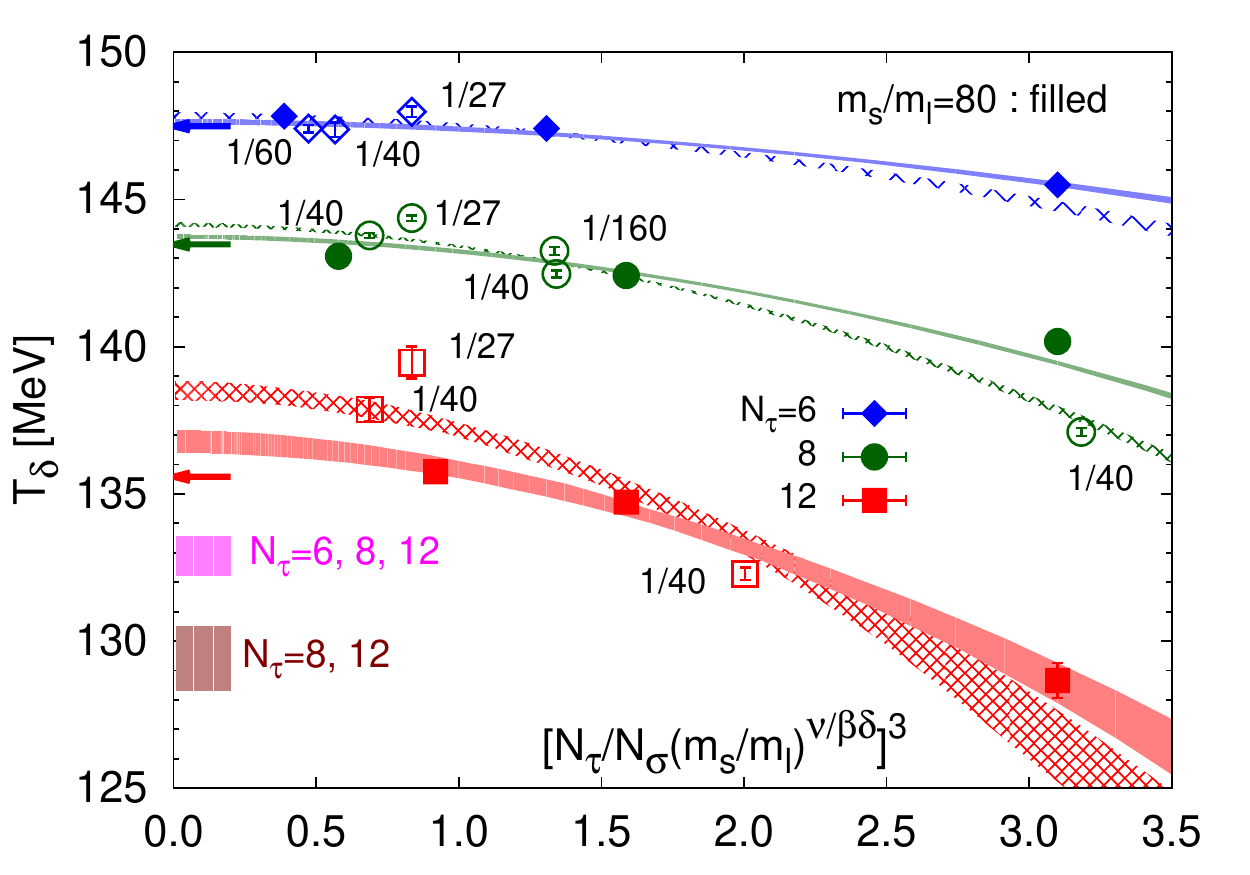}
\caption{{\it Left}: 
The ratio $H\chi_M/M$ versus temperature for $N_\tau=12$,
$m_l/m_s=1/80$ and different spatial volumes. {\it Middle}: 
Infinite-volume extrapolations based on an $O(4)$ finite-size scaling
ansatz (colored bands) and fits linear in $1/V$ (gray bands). Horizontal
bars show the continuum extrapolated results for $H=1/80$.
{\it Right}: Finite-size scaling fits for $T_\delta$ based on all
data for $H\le 1/27$ and all available volumes. Arrows show chiral limit
results at fixed $N_\tau$, and horizontal
bars show the continuum extrapolated results for $H=0$.
}
\label{fig:HchiM}
\end{figure*}

In Fig.~\ref{fig:sus}~(right), we show the volume dependence of $\chi_M$ for 
$H=1/80$ on lattices with temporal extent $N_\tau=8$ and for $N_\sigma/N_\tau=4,\ 5$, and $7$. 
Similar results have also been obtained for $N_\tau=6$ and $12$.
We note that $\chi_M^{\rm max}$ decreases slightly with increasing volume,
contrary to what one would expect to find at or close to a first- or
second-order phase transition. Our current results, thus, 
are consistent with a continuous phase transition at $H_c=0$.

\begin{figure}[h]
\includegraphics[width=0.5\textwidth]{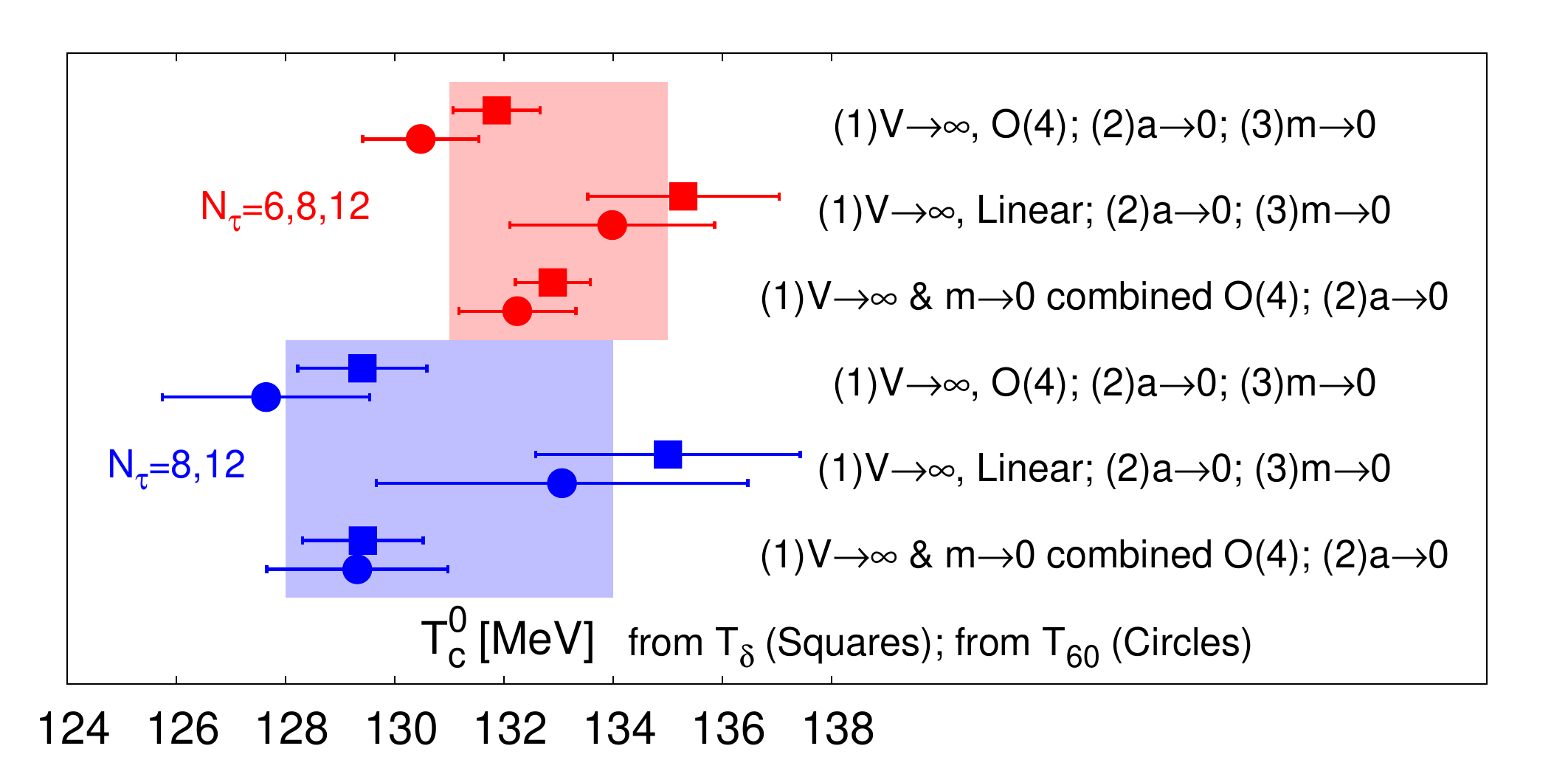}
\caption{Summary of fit results. For details, see text.}
\label{fig:final}
\end{figure}

Using results for $\chi_M$ and $M$ we constructed the
ratios $H\chi_M /M$ for different lattice sizes and several values of the
quark masses. This is shown in Fig.~\ref{fig:HchiM}~(left) for the lightest 
quark masses used on the $N_\tau=12$ lattices, $H=1/80$. The intercepts
with the horizontal line at $1/\delta$ define $T_\delta(H,L)$. 
For $H=1/80$ and each
of the three temporal lattice sizes
we have results for three different volumes on which we can extrapolate
$T_\delta(H,L)$ to the infinite-volume limit. We performed such extrapolations
using (i) the $O(4)$ ansatz given in Eq.~(\ref{TX}) as well as (ii) an 
extrapolation in $1/V$. The latter is appropriate for large $L$, if the 
volume dependence predominantly arises from regular terms and the former
is appropriate close to or in the continuum limit, if the singular part
dominates the partition function.
In the former case, we use the approximation $z_\delta(z_L) \sim z_L^{5.7}$,
which parametrizes well the finite-size dependence of $T_\delta$
in the scaling regime
\cite{Engels:2014bra}.
The resulting fits are shown in Fig.~\ref{fig:HchiM}~(middle). 
We note that results
for fixed $H$ tend to approach the infinite-volume limit more rapidly 
than $1/V$, which
is in accordance with the behavior expected from the ratio of finite-size
scaling functions. 
The resulting continuum limit extrapolations in $1/N_\tau^2$ based on
data for (i) all three $N_\tau$ values, as well as (ii) $N_\tau=8$ and $12$ 
only, are shown as horizontal bars in this figure. 
An analogous analysis is performed for $H=1/40$. 
Finally, we extrapolate the 
continuum results for $T_\delta(H,\infty)$ with $H=1/40$
and $1/80$ to the chiral limit using Eq.~(\ref{TX})
with $z_\delta(0)=0$.
Results obtained from these extrapolation chains, which involve
either a $1/V$ or $O(4)$ ansatz for the infinite-volume extrapolation,
as well as continuum limit extrapolations performed on
two different datasets, lead to chiral transition temperatures $T_c^0$ 
in the range ($128$-$135$)~MeV.
The resulting values for $T_c^0$ are summarized in Fig.~\ref{fig:final}.

As the fits shown in Fig.~\ref{fig:HchiM}~(middle) suggest that the
$O(4)$ scaling ansatz is appropriate for the analysis of finite-volume
effects already at nonzero values of the cutoff,
we can attempt 
a combined analysis of all data available for different light quark
masses and volumes at fixed $N_\tau$. This utilizes the quark mass dependence
of finite-size corrections, expressed in terms of $z_L$, and thus it intertwines
continuum and chiral limit extrapolations.~Using 
the scaling ansatz given in Eq.~(\ref{TX}), it also allows us to account 
for the contribution of a regular term in a single fit. 
Fits for fixed $N_\tau$ based on this ansatz, using data for all available 
lattice sizes and $H\le 1/27$, are shown in 
Fig.~\ref{fig:HchiM}~(right). For each $N_\tau$, the fit yields results for 
$T_\delta(H,L)$ at arbitrary $H$.
Some bands for $H=1/40$ and $1/80$ are shown in
the figure. As can be seen, for $H=1/80$, these bands compare well 
with the fits shown in Fig.~\ref{fig:HchiM}~(middle). 
For each $N_\tau$ an arrow shows
the corresponding chiral limit result, $T_\delta(0,\infty)$. 
We extrapolated these chiral limit
results to the continuum limit and estimated systematic errors again by
including or leaving out data for $N_\tau=6$. 
The resulting $T_c^0$ vaules, shown in Fig.~\ref{fig:final}, are in complete
agreement with the corresponding numbers obtained by first taking the
continuum limit and then taking the chiral limit.
Within the current accuracy these two limits are interchangeable. 

Similarly we analyzed results for $T_{60}$ on all datasets
using the same analysis strategy as for $T_\delta$. As can be 
seen in Fig.~\ref{fig:final}, we find for each extrapolation
ansatz that the resulting values for $T_c^0$ 
agree to within better than 1\% accuracy with the corresponding values 
for $T_\delta$.
This corroborates that
the chiral susceptibilities used for this analysis reflect basic features
of the $O(4)$ scaling functions. 

Performing continuum extrapolations by either including or discarding
results obtained on the coarsest ($N_\tau=6$) lattices leads to
a systematic shift of about $2$-$3$~MeV in the estimates for $T_c^0$.
This is reflected in the displacement of the two bands
in Fig.~\ref{fig:final}, which show averages for $T_c^0$ obtained with our 
different extrapolation {\it Ans\"atze}. 
Averaging separately over results for $T_\delta$ and $T_{60}$ 
obtained with both continuum extrapolation
procedures and including this systematic effect,
we find for the chiral phase transition temperature
\begin{equation}
T_c^0 = 132^{+3}_{-6}~{\rm MeV} \; .
\label{Tcfinal}
\end{equation}

\emph{Conclusions.---} Based on two novel estimators, we have determined the chiral
phase transition temperature in QCD with two massless light quarks and a
physical strange quark. Equation~(\ref{Tcfinal}) gives our thermodynamic-, continuum-, and
chiral- extrapolated  result for the chiral phase transition temperature,  which is
about $25$~MeV smaller than the pseudocritical (crossover)  temperature $T_{pc}$ for physical
values of the light and strange quark masses~\cite{Bazavov:2018mes}. Lattice QCD calculations presented here were carried out using the so-called ``rooted" staggered fermion formulation. There are ample theoretical and numerical evidences (for a review, see Ref. \cite{Sharpe:2006re}) that once the proper order of the limits---first continuum and then chiral---is followed, this formulation produces correct physical results~\cite{Bernard:2006vv,Bernard:2004ab}. In the present calculations we followed the proper order of the limits. However, we also checked that the quoted value $T_c^0$ remained unchanged, within our numerical accuracies, even when joint chiral and continuum limits were carried out. Notwithstanding such reassuring checks, in the future it will be important to carry out similar lattice QCD calculations using other fermion actions. The two estimators proposed in the current Letter will also be useful in such calculations.


\vspace{0.3cm}
This work was supported in part by
the Deutsche Forschungsgemeinschaft (DFG) through Grant No. 315477589-TRR 211,
by Grants No. 05P15PBCAA and No. 05P18PBCA1 of the German Bundesministerium f\"ur Bildung und
Forschung, and by the National Natural
Science Foundation of China under Grants No. 11775096 and No. 11535012. Furthermore, this work was supported by Contract No.~DE-SC0012704 with the
U.S. Department of Energy, by the Scientific Discovery through Advanced
Computing (SciDAC) program funded by the U.S. Department of Energy, by the Office of
Science, Advanced Scientific Computing Research and Nuclear Physics, by
the DOE Office of Nuclear Physics funded BEST topical collaboration, 
and by a Early Career Research Award of the Science and Engineering
Research Board of the Government of India.
Numerical calculations have been made possible through PRACE grants
at CSCS, Switzerland, and at CINECA, Italy as well as grants at the
Gauss Centre for Supercomputing and 
NIC-J\"ulich, Germany. These grants provided access to resources on
Piz Daint at CSCS and Marconi at CINECA, as well as on 
JUQUEEN and JUWELS at NIC.
Additional calculations have been performed on 
GPU clusters of USQCD, 
at Bielefeld University, the PC$^2$ Paderborn 
University, and the Nuclear Science Computing Center at Central China
Normal University (NSC$^3$), Wuhan, China. Some datasets have also partly been
produced at the TianHe II Supercomputing Center in Guangzhou.


\end{document}